\documentclass[12pt,prd,onecolumn,showpacs,amsmath,amssymb,aps,floats,floatfix,nofootinbib]{revtex4-1}

\usepackage[colorlinks=true,urlcolor=blue,anchorcolor=blue,citecolor=blue,filecolor=blue,linkcolor=blue,menucolor=blue,linktocpage=true]{hyperref} 

\usepackage[inline]{enumitem}
\usepackage{dcolumn}
\usepackage{bm}
\usepackage{amsmath}
\usepackage{amsfonts}
\usepackage{amssymb}
\usepackage{color}
\usepackage{float}
\usepackage{latexsym}
\usepackage{slashed} 
\usepackage{pstricks}
\usepackage{indentfirst}
\usepackage{mathrsfs}
\usepackage{multirow}
\usepackage{epsfig,psfrag}
\usepackage{subfigure}
\usepackage{mathtools}
\usepackage{setspace} 
\usepackage{lipsum}
\usepackage[utf8]{inputenc} 
\usepackage{siunitx}
\usepackage[normalem]{ulem}
\usepackage{booktabs}
\usepackage{numprint}

\graphicspath{{fig/}}

\usepackage{array}
\newcommand{\PreserveBackslash}[1]{\let\temp=\\#1\let\\=\temp}
\newcolumntype{C}[1]{>{\PreserveBackslash\centering}p{#1}}
\newcolumntype{R}[1]{>{\PreserveBackslash\raggedleft}p{#1}}
\newcolumntype{L}[1]{>{\PreserveBackslash\raggedright}p{#1}}

\newcommand{\NGG}{{\mathrm{SU(2)_{D}}}}
\newcommand{\gd}{g_{\mathrm{D}}}

\newcommand{\Lag}{\mathcal{L}}
\newcommand{\deli}[3]{\left#1 #2 \right#3}
\newcommand{\SD}[1]{\deli{(}{#1}{)}}

\newcommand{\sv}{\langle\sigma_{\mathrm{ann}}v\rangle}
\newcommand{\SUtwoL}{\mathrm{SU}(2)_\mathrm{L}}
\newcommand{\UoneY}{\mathrm{U}(1)_\mathrm{Y}}
\newcommand{\Uone}{\mathrm{U}(1)}

\let\isout\sout
\renewcommand\sout{\bgroup\markoverwith {\textcolor{blue}{\rule[0.5ex]{2pt}{0.4pt}}}\ULon}
\let\isout\sout
\renewcommand{\sout}[1]{\ifmmode\text{\isout{\ensuremath{#1}}}\else\isout{#1}\fi}

\allowdisplaybreaks 

\begin{document}

\title{Vector dark matter from split SU(2) gauge bosons}
\author{Zexi Hu}
\author{Chengfeng Cai}
\author{Yi-Lei Tang}\email[Corresponding author. ]{tangylei@mail.sysu.edu.cn}
\author{Zhao-Huan Yu}\email[Corresponding author. ]{yuzhaoh5@mail.sysu.edu.cn}
\author{Hong-Hao Zhang}\email[Corresponding author. ]{zhh98@mail.sysu.edu.cn}
\affiliation{School of Physics, Sun Yat-Sen University, Guangzhou 510275, China}

\begin{abstract}

We propose a vector dark matter model with an exotic dark $\mathrm{SU}(2)$ gauge group. Two Higgs triplets are introduced to spontaneously break the symmetry. All of the dark gauge bosons become massive, and the lightest one is a viable vector DM candidate. Its stability is guaranteed by a remaining $Z_2$ symmetry. We study the parameter space constrained by the Higgs measurement data, the dark matter relic density, and direct and indirect detection experiments. We find numerous parameter points satisfying all the constraints, and they could be further tested in future experiments. Similar methodology can be used to construct vector dark matter models from an arbitrary $\mathrm{SO}(N)$ gauge group.

\end{abstract}

\maketitle
\tableofcontents

\section{Introduction}
\label{sec:intro}

Dark matter (DM) occupies approximately 26\% of the energy density of the present Universe.
In the thermal paradigm, DM particles annihilate into standard model (SM) products in the early Universe before they freeze out from the thermal plasma, leaving a proper relic density~\cite{Bertone:2004pz,Feng:2010gw,Young:2016ala}.
Particle candidates for dark matter can be classified according to their spins.
Thus, DM particles could be scalar bosons (spin-0), spin-1/2 fermions, vector bosons (spin-1), or spin-3/2 fermions.
Various scalar and fermionic DM models have been adequately studied in the literature.
On the other hand, vector DM models draw much less attentions, with many issues unstudied.

When the related mediators are sufficiently heavy, interactions of vector DM can be appropriately described by effective operators in a model-independent way~\cite{Belanger:2008sj, Yu:2011by}.
Otherwise, renormalizable interactions should be considered.
If extra dimensions exist, the first Kaluza-Klein mode of the $\UoneY$ gauge boson could be a well motivated vector DM candidate~\cite{Servant:2002aq, Cheng:2002ej}.
In the four-dimensional spacetime, a natural approach for constructing renormalizable vector DM models utilizes the gauge theories, in which at least one gauge boson acts as the DM particle.
A mechanism is required to generate the mass for the vector DM particle.
For $\Uone$ gauge theories, this mechanism can be either the Stueckelberg mechanism~\cite{Stueckelberg:1900zz, Chodos:1971yj} or the Brout-Englert-Higgs mechanism~\cite{Higgs:1964ia, Higgs:1964pj, Englert:1964et}.
For non-abelian gauge theories, the former is inapplicable, and the latter is commonly considered.
Furthermore, the DM particle could be a confined spin-1 bound state based on non-abelian gauge interactions, and thus its mass is linked to a confinement scale~\cite{Hambye:2009fg}.

Focusing on fundamental DM particles, a simple vector DM model can be constructed by introducing a dark $\mathrm{U(1)}$ gauge group with a dark Higgs field providing the gauge boson mass and a portal to the SM sectors~\cite{Lebedev:2011iq, Abe:2012hb, Farzan:2012hh, Baek:2012se, Domingo:2013tna, Yu:2014pra, Chen:2014cbt, Gross:2015cwa, Duch:2015jta, DiFranzo:2015nli, Azevedo:2018oxv, Mohamadnejad:2019vzg, Glaus:2019itb, Arcadi:2020jqf, Delaunay:2020vdb, Salehian:2020asa}.
A $Z_2$ symmetry is required to guarantee the stability of the vector DM candidate.
More complicated extensions involve non-abelian gauge groups, such as
$\mathrm{SU(2)}$~\cite{Hambye:2008bq, Baek:2013dwa, Boehm:2014bia, Khoze:2014woa, Chen:2015nea, Gross:2015cwa, DiChiara:2015bua, Chen:2015dea, Karam:2015jta, Khoze:2016zfi, Barman:2018esi, Saez:2018off, Belyaev:2018xpf, Barman:2019lvm, Ghosh:2020ipy},
$\mathrm{SU(2)} \otimes \mathrm{U(1)}$~\cite{Chiang:2013kqa, Davoudiasl:2013jma, Choi:2019zeb, Ramos:2021omo},
$\mathrm{SU(2)} \otimes \mathrm{SU(2)}$~\cite{Abe:2020mph},
$\mathrm{SU(3)}$~\cite{Gross:2015cwa,  DiChiara:2015bua, Karam:2016rsz, Ko:2016fcd, Arcadi:2016kmk, Poulin:2018kap}, and general $\mathrm{SU}(N)$~\cite{Gross:2015cwa, DiChiara:2015bua}.
Typically, some discrete or global continuous symmetries remain after gauge symmetry breaking, stabilizing the vector DM particle.

If the dark gauge group is $\mathrm{SU(2)}$, it can be spontaneously broken completely through one dark Higgs doublet, with a remaining custodial global $\mathrm{SU(2)}$ symmetry ensuring the stability of the three degenerate gauge bosons as vector DM particles~\cite{Hambye:2008bq}.
If, instead, one real Higgs triplet is introduced to break the $\mathrm{SU(2)}$ gauge symmetry, a $\mathrm{U(1)}$  gauge symmetry would remain, leading to a massless gauge boson acting as the dark radiation~\cite{Baek:2013dwa, Ghosh:2020ipy}.
The other two gauge bosons are massive and degenerate, forming a pair of vector DM particle and antiparticle.
This scenario can also be cast to a dark $\mathrm{SU(3)}$ case, where one Higgs triplet partially breaks the gauge group, leaving us five DM vector bosons and three massless dark radiation particles~\cite{Ko:2016fcd}.
On the other hand, two dark Higgs triplets would be able to generate masses for all the eight $\mathrm{SU(3)}$ gauge bosons~\cite{Gross:2015cwa, Arcadi:2016kmk, Poulin:2018kap}.
For a general dark $\mathrm{SU}(N)$ gauge group, all the gauge bosons can be made massive if $N-1$ Higgs fields in the fundamental representation are introduced~\cite{Gross:2015cwa}.

In this paper, we study a vector DM model with a dark $\mathrm{SU(2)}_\mathrm{D}$ gauge group broken by two real Higgs triplets, which develop a generic configuration of vacuum expectation values (VEVs).
As a result, the three dark gauge bosons can obtain different masses.
The two lighter gauge bosons are odd under a remaining $Z_2$ symmetry, and the lightest one is stable, playing the role of the DM candidate.
Compared with Ref.~\cite{Hambye:2008bq, Baek:2013dwa, Ghosh:2020ipy}, no mass degeneracy appears among the three dark gauge bosons.
After spontaneous symmetry breaking, the dark Higgs bosons mix with the $\mathrm{SU(2)}_\mathrm{L}$ Higgs boson, and there are four mass eigenstates of neutral Higgs bosons.
The dark sector communicates with the SM sectors only through the Higgs quartic couplings.
Thus, this model belongs to a kind of Higgs-portal DM models.
We will perform a random scan in the parameter space to investigate phenomenological constraints from collider measurements of the 125~GeV Higgs boson, direct collider searches for exotic Higgs bosons, electroweak precision measurements, the DM relic density, and the direct/indirect DM detection experiments.
Finally, we will discuss the possibility of the generalization to dark $\mathrm{SO(N)}$ gauge groups for $N>2$ with $N-1$ real Higgs multiplets in the $N$-dimensional fundamental representation.
We will prove that an accidental $Z_2$ symmetry also arises to preserve the stability of a dark gauge boson, acting as a DM candidate.

In Sec.~\ref{sec:model}, we describe the model setup and discuss the general formalism of the remaining $Z_2$ symmetry.  In Sec.~\ref{sec:pheno}, the constraints from the relic density and the direct and indirect detection experiments are displayed.
In Sec.~\ref{sec:extension}, we generalize the methodology to a general dark $\mathrm{SO}(N)$ model.
In Sec.~\ref{sec:conclusion}, we summarize the paper.

\section{Model}
\label{sec:model}

Now we discuss our vector DM model based on the dark $\mathrm{SU(2)}_\mathrm{D}$ gauge symmetry.
The corresponding gauge fields are denoted as $\tilde{A}^a_{\mu}$ ($a=1,2,3$).
Two real $\mathrm{SU(2)}_\mathrm{D}$ Higgs triplets, $\Phi_a$ and $X_a$ ($a=1,2,3$), are introduced to break the $\mathrm{SU(2)}_\mathrm{D}$ gauge symmetry.
We call them dark Higgs fields, as they are assumed to be SM gauge singlets.
In this section, we construct the generic Lagrangian, and study the spontaneous symmetry breaking in detail.

\subsection{Lagrangian and symmetry analysis}

Under a $\mathrm{SU(2)}_\mathrm{D}$ gauge transformation, the dark Higgs fields $\Phi_a$ and $X_a$ transform as
\begin{equation}
\Phi \to U \Phi,\quad
X \to U X,
\end{equation}
where $U = \exp[i \theta^{a} (x)\mathcal{T}^{a}]$ with $(\mathcal{T}^{a})_{bc}$ being the generators in the adjoint representation of $\mathrm{SU(2)}_\mathrm{D}$, and some representation indices have been omitted.
Below, we adopt $(\mathcal{T}^{a})_{bc} = -i \varepsilon^{abc}$, where $\varepsilon^{abc}$ is the three-dimensional Levi-Civita symbol.
The corresponding transformation of the $\tilde{A}^a_{\mu}$ is
\begin{equation}
\tilde{A}^a_{\mu} \mathcal{T}^{a} \to U \tilde{A}^a_{\mu} \mathcal{T}^{a} U^{\dag} + \frac{i}{\gd} U \partial_{\mu} U^{\dag} ,
\end{equation}
where $\gd$ is the $\mathrm{SU(2)}_\mathrm{D}$ gauge coupling.

$\Phi_a$ and $X_a$ can be treated as vectors in a three-dimensional Euclidean representation space.
From three vectors $\mathbf{E}$, $\mathbf{F}$, and $\mathbf{G}$ in such a space,  nonzero invariant scalars  can only be constructed via either dot products, like $\mathbf{E} \cdot \mathbf{F} = E_a F_a$ and $\mathbf{G} \cdot \mathbf{G} = G_a G_a$, or triple products, like $(\mathbf{E} \times \mathbf{F}) \cdot \mathbf{G} = \varepsilon^{abc} E_a F_b G_c$.
Since we only have two triplet Higgs fields $\Phi_a$ and $X_a$, the related scalar potential terms must be formed by bilinear dot products $\Phi_a \Phi_a$, $X_a X_a$, and $\Phi_a X_a$.
Therefore, the model respects an accidental $Z_2$ symmetry, under which the Higgs triplets transform as
\begin{equation}
\Phi \rightarrow P_\mathrm{D} \Phi = -\Phi,\quad
X \rightarrow P_\mathrm{D} X = -X,
\end{equation}
where $P_\mathrm{D} = \operatorname{diag}(-1, -1, -1)$ serves as a ``dark parity'', analogous to the parity $P$ in the 3-dimensional physical space.
Thus, $\Phi_a$ and $X_a$ are $P_\mathrm{D}$-odd, while the rest fields are $P_\mathrm{D}$-even.
The $Z_2$ symmetry and the global $\NGG \simeq \mathrm{SO(3)_D}$ symmetry form a global $\mathrm{O(3)_D}$ symmetry, with $P_\mathrm{D} \in \mathrm{O(3)_D}$.
For any element matrix $R \in \mathrm{O(3)_D}$, the transformations of the $\mathrm{O(3)_D}$ vectors $\Phi_a$ and $X_a$ are given by
\begin{equation}\label{eq:O3_trans:Phi_X}
\Phi_a \rightarrow R_{ab} \Phi_b,\quad
X_a \rightarrow R_{ab} X_b.
\end{equation}

The generic renormalizable scalar potential reads
\begin{equation}
V = V_{\mathrm{SM}} + V_{\mathrm{D}} + V_{\mathrm{P}} ,
\end{equation}
where the SM part for the $\SUtwoL$ Higgs doublet $H$ is
\begin{equation}
V_{\mathrm{SM}} = - \mu_{0}^{2} |H|^{2} + \lambda_{0} |H|^{4} ,
\end{equation}
the dark sector part is
\begin{eqnarray}
  V_{\mathrm{D}} &=& - \mu_{1}^{2}\Phi_{a}\Phi_{a} - \mu_{2}^{2}X_{a}X_{a} - \mu_{3}^{2}\Phi_{a}X_{a} + \lambda_{1}(\Phi_{a}\Phi_{a})^2 + \lambda_{2}(X_{a}X_{a})^2
\nonumber\\
  && + \lambda_{3}\Phi_{a}\Phi_{a}X_{b}X_{b} + \lambda_{4}\Phi_{a}\Phi_{a}\Phi_{b}X_{b} + \lambda_{5}\Phi_{a}X_{a}X_{b}X_{b} + \lambda_{6}(\Phi_{a}X_{a})^2 ,
\end{eqnarray}
and the portal part is
\begin{eqnarray}
V_{\mathrm{P}} = \lambda_{10}|H|^{2}\Phi_{a}\Phi_{a} + \lambda_{20}|H|^{2}X_{a}X_{a} + \lambda_{30}|H|^{2}\Phi_{a}X_{a}.
\end{eqnarray}
It is easy to verify that $V$ is invariant under the global $\mathrm{O(3)_D}$ transformations.

The Lagrangian of the model can be divided into
\begin{equation}
  \label{eq: Model Lagrangian}
  \Lag = \Lag_{\mathrm{SM}} + \Lag_{\mathrm{D}} ,
\end{equation}
where $\Lag_{\mathrm{SM}}$ is the SM Lagrangian involving $-V_\mathrm{SM}$, and
\begin{equation}
  \Lag_{\mathrm{D}} = - \frac{1}{4} \tilde A^{a}_{\mu\nu}\tilde A^{a,\mu\nu} + \frac{1}{2}(D_{\mu}\Phi_{a})^T(D^{\mu}\Phi_{a}) + \frac{1}{2}(D_{\mu}X_{a})^T(D^{\mu}X_{a}) - V_{\mathrm{D}} - V_{\mathrm{P}}
\end{equation}
is the Lagrangian for the dark sector with $\tilde A^{a}_{\mu\nu} \equiv {\partial _\mu }\tilde A_\nu ^a - {\partial _\nu }\tilde A_\mu ^a + {g_{\mathrm{D}}}{\varepsilon ^{abc}}{{\tilde A}^{b,\mu }}{{\tilde A}^{c,\nu }}$.
The covariant derivatives of the dark Higgs triplets are
\begin{eqnarray}
{D_\mu }{\Phi _a} &=& {\partial _\mu }{\Phi _a} - i{g_{\mathrm{D}}}\tilde A_\mu ^c{({\mathcal{T}^c})_{ab}}{\Phi _b}
= {\partial _\mu }{\Phi _a} + {g_{\mathrm{D}}}{\varepsilon ^{acb}}\tilde A_\mu ^c{\Phi _b},
\\
{D_\mu }{X _a} &=& {\partial _\mu }{X _a} - i{g_{\mathrm{D}}}\tilde A_\mu ^c{({\mathcal{T}^c})_{ab}}{X _{b}}
= {\partial _\mu }{X _a} + {g_{\mathrm{D}}} {\varepsilon ^{acb}} \tilde A_\mu ^c{X _b}.
\end{eqnarray}
$\Lag_{\mathrm{D}}$ involves gauge interaction terms in a triple-product form, e.g., $\gd\varepsilon^{abc}\tilde{A}^{a}_{\mu}\Phi_{b}\partial^{\mu}\Phi_{c}$.
Since the Levi-Civita symbol satisfies $\varepsilon^{abc} = \det (R) R_{ad} R_{be} R_{cf} \varepsilon^{def}$ for $R \in \mathrm{O(3)_D}$, the $\mathrm{O(3)_D}$ invariance of the Lagrangian requires that $\tilde{A}^a_\mu$ acts as an $\mathrm{O(3)_D}$ axial vector, whose transformation is
\begin{equation}\label{eq:O3_trans:A}
\tilde{A}^a_\mu \to \det(R) R_{ab} \tilde{A}^b_\mu,\quad
R \in \mathrm{O(3)_D}.
\end{equation}
Thus, the $P_\mathrm{D}$ transformation of $\tilde{A}^a_\mu$ is
\begin{eqnarray}
\tilde{A}^a_\mu \to \det(P_\mathrm{D}) P_{\mathrm{D},ab} \tilde{A}^b_\mu = + \tilde{A}^a_\mu,
\end{eqnarray}
i.e., $\tilde{A}^a_\mu$ is $P_\mathrm{D}$-even.

\subsection{Spontaneous symmetry breaking}
\label{subsec:spon_sym_break}

In order to generate masses for the dark gauge bosons, the two real dark Higgs triplets $\Phi_a$ and  $X_a$ should obtain VEVs $\langle \Phi_a \rangle$ and $\langle X_a \rangle$, respectively.
Thus, the $\NGG$ gauge symmetry is spontaneously broken, and so is the global $\mathrm{O(3)_D}$ symmetry.
In general, the vectors $\langle \Phi_a \rangle$ and $\langle X_a \rangle$ are not parallel to each other, so they determine a plane in the three-dimensional representation space.
We can always rotate the axes to a configuration that the $z$-axis is along the $\langle \Phi_a \rangle$ direction and the $y$-axis lies inside the plane.
Therefore, without loss of generality, the configuration of the VEVs can be expressed as $\langle \Phi_a \rangle = (0,0,v_1)$ and $\langle X_a \rangle = (0,v_2,v_3)$.
Expanding the fields around the VEVs, we have
\begin{equation}
\Phi = \begin{pmatrix}
\phi_{1}\\
\phi_{2}\\
v_{1} + \phi_{3}
\end{pmatrix},\quad
X = \begin{pmatrix}
\chi_{1}\\
v_{2} + \chi_{2}\\
v_{3} + \chi_{3}
\end{pmatrix}.
\label{vevConvention}
\end{equation}
In the SM sector, the $\SUtwoL$ Higgs doublet $H$ has a form of
\begin{equation}
  H = \frac{1}{\sqrt{2}}
    \begin{pmatrix}
        0\\
        v_0 + \tilde{h}_{0}
    \end{pmatrix}
\end{equation}
in the unitary gauge.
Note that all the four VEVs $v_0$, $v_1$, $v_2$, and $v_3$ are real constants.

Expanding the covariant kinetic terms of the dark Higgs triplets, we have
\begin{eqnarray}
 \mathcal{L}_\mathrm{D} &\supset & \frac{1}{2}\SD{\partial_{\mu}\phi_{a}}^{2}
 + \frac{1}{2}\SD{\partial_{\mu}\chi_{a}}^{2}
 + \frac{1}{2} \mathcal{M}^{2}_{A,ab} \tilde{A}^{a}_{\mu}\tilde{A}^{b,\mu} + \gd\varepsilon^{abc}\tilde{A}^{a}_{\mu}(\phi_{b}\partial^{\mu}\phi_{c}
 + \chi_{b}\partial^{\mu}\chi_{c} )
\nonumber\\
  &&- \gd \tilde{A}^{1}_{\mu}\partial^{\mu}\SD{v_{1}\phi_{2} - v_{2}\chi_{3} + v_{3}\chi_{2}} + \gd \tilde{A}^{2}_{\mu}\partial^{\mu}\SD{v_{1}\phi_{1} + v_{3}\chi_{1}} - \gd v_{2}\tilde{A}^{3}_{\mu}\partial^{\mu}\chi_{1}
\nonumber\\
  && -\gd^{2}\big[v_{1}
(\tilde{A}^{3}_{\mu}\tilde{A}^{1,\mu}\phi_{1}
 + \tilde{A}^{3}_{\mu}\tilde{A}^{2,\mu}\phi_{2}
 - \tilde{A}^{1}_{\mu}\tilde{A}^{1,\mu}\phi_{3}
 - \tilde{A}^{2}_{\mu}\tilde{A}^{2,\mu}\phi_{3})
\nonumber\\
  && \quad\quad~ + v_2 (\tilde{A}^{2}_{\mu}\tilde{A}^{1,\mu}\chi_{1}
 + \tilde{A}^{2}_{\mu}\tilde{A}^{3,\mu}\chi_{3}
 - \tilde{A}^{1}_{\mu}\tilde{A}^{1,\mu}\chi_{2}
 - \tilde{A}^{3}_{\mu}\tilde{A}^{3,\mu}\chi_{2})
\nonumber\\
  && \quad\quad~ + v_3
(\tilde{A}^{3}_{\mu}\tilde{A}^{1,\mu}\chi_{1}
 + \tilde{A}^{3}_{\mu}\tilde{A}^{2,\mu}\chi_{2}
 - \tilde{A}^{1}_{\mu}\tilde{A}^{1,\mu}\chi_{3}
 - \tilde{A}^{2}_{\mu}\tilde{A}^{2,\mu}\chi_{3})\big]
\nonumber\\
  && - \frac{1}{2}\gd^{2}\sum\limits_{a \neq b}\big[\tilde{A}^{a}_{\mu}\tilde{A}^{b,\mu}(\phi_{a}\phi_{b} + \chi_{a}\chi_{b}) - \tilde{A}^{a}_{\mu}\tilde{A}^{a,\mu}(\phi_{b}\phi_{b} + \chi_{b}\chi_{b})\big],
\label{eq: Covariant kinetic terms}
\end{eqnarray}
where the mass-squared matrix for the dark gauge bosons is
\begin{equation}
  \mathcal{M}_{A}^{2} = \gd^{2}
  \begin{pmatrix}
    v_{123}^2 & 0 & 0 \\
    0 & v_{13}^2 & -v_{2}v_{3} \\
    0 & -v_{2}v_{3} & v_{2}^{2}
  \end{pmatrix}. \label{AMass2}
\end{equation}
For convenience, we define
\begin{equation}
v_{123} \equiv \sqrt{v_1^2 + v_2^2 + v_3^2},\quad
v_{13} \equiv \sqrt{v_1^2 + v_3^2},\quad
v_{23} \equiv \sqrt{v_2^2 + v_3^2}.
\end{equation}

$\mathcal{M}_{A}^{2}$ can be diagonalized as
\begin{equation}
\mathcal{O}_{A}^\mathrm{T} \mathcal{M}_{A}^{2} \mathcal{O}_{A} = \operatorname{diag} (m_{A^1}^2,\, m_{A^2}^2,\, m_{A^3}^2)
\end{equation} by an orthogonal matrix
\begin{equation}
\mathcal{O}_{A} =  \begin{pmatrix}
    1 & 0 & 0 \\
    0 & \cos\theta & -\sin\theta \\
    0 & \sin\theta & \cos\theta
  \end{pmatrix}
\end{equation}
with
\begin{equation}
  \sin\theta = \sqrt{2} v_2 v_3 \left[ v_{123}^4 - 4 v_1^2 v_2^2 + (v_2^2 - v_1^2 - v_3^2) \sqrt{v_{123}^4 - 4 v_1^2 v_2^2} \right]^{-1/2} .
\end{equation}
The relation between $\tilde{A}_{\mu}^{a}$ and the mass eigenstates $A_{\mu}^a$ is
\begin{equation}
\tilde{A}_{\mu}^{a} = \mathcal{O}_{A,ab} A^{b}_{\mu} .
\end{equation}
Note that $A^1_\mu = \tilde{A}^1_\mu$, while $A^2_\mu$ and $A^3_\mu$  are linear combinations of $\tilde{A}^2_\mu$ and $\tilde{A}^3_\mu$.
The physical masses squared are given by
\begin{eqnarray}
  \label{eq: A2 A2 masses}
    m_{A^1}^2 &=& g_\mathrm{D}^2 v_{123}^2 , \\
    m_{A^2}^2 &=& \frac{\gd^2}{2} \left(v_{123}^2 - \sqrt{v_{123}^4 - 4 v_1^2 v_2^2}\right) , \\
    m_{A^3}^2 &=& \frac{\gd^2}{2} \left(v_{123}^2 + \sqrt{v_{123}^4 - 4 v_1^2 v_2^2}\right) .
\end{eqnarray}
Thus, we find $m_{A^2} \leq m_{A^3} \leq m_{A^1}$.
For the generic nonzero $v_1$, $v_2$, and $v_3$, there is no degeneracy in the mass eigenstates.

We further identify the Goldstone bosons
\begin{eqnarray}
  G_{1} &=& v_{123}^{-1} (v_{1}\phi_{2} + v_{3}\chi_{2} - v_{2}\chi_{3}) ,
\\
  G_{2} &=& {(c_\theta ^2v_1^2 + s_\theta ^2v_2^2 + c_\theta ^2v_3^2)^{ - 1/2}}[ - {c_\theta }{v_1}{\phi _1} + ({s_\theta }{v_2} - {c_\theta }{v_3}){\chi _1}] ,
\\
  G_{3} &=& {(s_\theta ^2v_1^2 + c_\theta ^2v_2^2 + s_\theta ^2v_3^2)^{ - 1/2}}[{s_\theta }{v_1}{\phi _1} + ({c_\theta }{v_2} + {s_\theta }{v_3}){\chi _1}] ,
\end{eqnarray}
which are eaten by the gauge bosons $A^1$, $A^2$, and $A^3$, respectively.
The shorthand notations $s_\theta \equiv \sin\theta$ and $c_\theta \equiv \cos\theta$ have been used.
The Higgs bosons orthogonal to these Goldstone bosons can be chosen as
\begin{eqnarray}
    \tilde{h}_{1} &=& v_{23}^{-1} (v_{2}\chi_{2} + v_{3}\chi_{3}),
\\
    \tilde{h}_{2} &=& (v_{23}v_{123})^{-1} (v_{23}^{2}\phi_{2} - v_{1}v_{3}\chi_{2} + v_{1}v_{2}\chi_{3}),
\\
    \tilde{h}_{3} &=& \phi_{3} .
\end{eqnarray}
Utilizing the minimization conditions for the potential $V$, we obtain
\begin{eqnarray}
\mu _0^2 &=& {\lambda _0}v_0^2 + {\lambda _{10}}v_1^2 + {\lambda _{20}}v_{23}^2 + {\lambda _{30}}{v_1}{v_3},
\\
\mu _1^2 &=& 2{\lambda _1}v_1^2 + {\lambda _3}v_{23}^2 + {\lambda _4}{v_1}{v_3} + \frac{1}{2}{\lambda _{10}}v_0^2,
\\
\mu _2^2 &=& 2{\lambda _2}v_{23}^2 + {\lambda _3}v_1^2 + {\lambda _5}{v_1}{v_3} + \frac{1}{2}{\lambda _{20}}v_0^2,
\\
\mu _3^2 &=& {\lambda _4}v_1^2 + {\lambda _5}v_{23}^2 + 2{\lambda _6}{v_1}{v_3} + \frac{1}{2}{\lambda _{30}}v_0^2.
\end{eqnarray}
We then derive the mass-squared matrix $\mathcal{M}^2_{h}$ for the Higgs bosons $(\tilde{h}_0, \tilde{h}_1, \tilde{h}_2, \tilde{h}_3)$, whose elements are given by
\begin{eqnarray}
\mathcal{M}^2_{h,00} &=& 2 \lambda_{0}v_{0}^{2},\quad
\mathcal{M}^2_{h,11} = 8 \lambda_{2}v_{23}^{2} + 4\lambda_{5}v_{1}v_{3} + 2\lambda_{6} {v_{1}^{2}v_{3}^{2}}/{v_{23}^{2}},
\\
\mathcal{M}^2_{h,22} &=& {2\lambda_{6}v_{2}^{2}v_{123}^{2}}/{v_{23}^{2}},\quad
\mathcal{M}^2_{h,33} = 8\lambda_{1}v_{1}^{2} + 4\lambda_{4}v_{1}v_{3} + 2\lambda_{6}v_{3}^{2},
\\
\mathcal{M}^2_{h,01} &=& 2\lambda_{20}v_{0}v_{23} + {\lambda_{30}v_{0}v_{1}v_{3}}/{v_{23}}, \quad
  \mathcal{M}^2_{h,02} = {\lambda_{30}v_{0}v_{2}v_{123}}/{v_{23}},
\\
\mathcal{M}^2_{h,03} &=& 2\lambda_{10}v_{0}v_{1} + \lambda_{30}v_{0}v_{3}, \quad
  \mathcal{M}^2_{h,12} = 2\lambda_{5}v_{2}v_{123} + {2\lambda_{6}v_{1}v_{2}v_{3}v_{123}}/{v_{23}^{2}},
\\
\mathcal{M}^2_{h,13} &=& 4\lambda_{3}v_{1}v_{23} + {2\lambda_{4}v_{1}^{2}v_{3}}/{v_{23}} + 2\lambda_{5}v_{3}v_{23} + {2\lambda_{6}v_{1}v_{3}^{2}}/{v_{23}},
\\
\mathcal{M}^2_{h,23} &=& {2\lambda_{4}v_{1}v_{2}v_{123}}/{v_{23}} + {2\lambda_{6}v_{2}v_{3}v_{123}}/{v_{23}}.
\end{eqnarray}

An orthogonal matrix $\mathcal{O}_{h}$ is used to diagonalize $\mathcal{M}_h^2$, leading to
\begin{equation}
\mathcal{O}_{h}^\mathrm{T} \mathcal{M}_h^2 \mathcal{O}_{h} = \operatorname{diag}(m_{h_0}^2, m_{h_1}^2, m_{h_2}^2, m_{h_3}^2).
\end{equation}
The basis $(\tilde{h}_0, \tilde{h}_1, \tilde{h}_2, \tilde{h}_3)$ can be rotated to the mass eigenstate basis $(h_0, h_1, h_2, h_3)$ through
\begin{equation}
\tilde{h}_{i} = \mathcal{O}_{h,ij} h_{j} .
\end{equation}
The elements of $\mathcal{O}_{h}$ will be numerically calculated.
We define $h_{0}$ as the SM-like Higgs boson.
In the following parameter scan, we require that the $h_{0}$ mass is around $125~\si{GeV}$ and $\tilde{h}_0$ makes the most contribution to $h_0$.
For the other Higgs bosons, we adopt a mass hierarchy convention of $m_{h_1} \leq m_{h_2} \leq m_{h_3}$.

Finally, we discuss a remaining $Z'_2$ symmetry after the spontaneous symmetry breaking of $\mathrm{O(3)_D}$.
This is the reflection symmetry with respect to the $\langle \Phi_a \rangle$-$\langle X_a \rangle$ plane, i.e., the $y$-$z$ plane according to our convention \eqref{vevConvention}, because the VEV configuration $\langle \Phi_a \rangle = (0,0,v_1)$ and $\langle X_a \rangle = (0,v_2,v_3)$ is preserved under such a reflection.
The corresponding reflection matrix $P_\mathrm{D}^{\prime} \in \mathrm{O(3)_D}$ is given by
\begin{eqnarray}
P_\mathrm{D}^{\prime} = P_\mathrm{D} e^{-i \pi \mathcal{T}^1} = \begin{pmatrix}
-1 & & \\
 & -1 & \\
 & & -1
\end{pmatrix}
\begin{pmatrix}
1 & & \\
 & -1 & \\
 & & -1
\end{pmatrix}
 = \begin{pmatrix}
-1 & & \\
 & 1 & \\
 & & 1
\end{pmatrix},
\end{eqnarray}
where $\SD{\mathcal{T}^1}_{bc} = -i \varepsilon^{1bc}$ is the rotation generator about the $x$-axis.
According to \eqref{eq:O3_trans:Phi_X} and \eqref{eq:O3_trans:A},
$\Phi_a$, $X_a$, and $\tilde A^a_\mu$ transform under the reflection as
\begin{eqnarray}
\Phi_a \rightarrow P_{\mathrm{D},ab}^{\prime} \Phi_b, \quad
X_a \rightarrow P_{\mathrm{D},ab}^{\prime} X_b, \quad
\tilde{A}^a_{\mu} \to \det(P_\mathrm{D}^{\prime}) P_{\mathrm{D},ab}^{\prime} \tilde{A}^b_{\mu} = -P_{\mathrm{D},ab}^{\prime} \tilde{A}^b_{\mu}.
\end{eqnarray}
Thus, we obtain
\begin{equation}
\phi_{1} \to -\phi_{1} ,\quad
\chi_{1} \to -\chi_{1} ,\quad
\tilde{A}^{2}_{\mu} \to - \tilde{A}^{2}_{\mu},\quad
\tilde{A}^{3}_{\mu} \to -\tilde{A}^{3}_{\mu}.
\end{equation}
These are the $P'_\mathrm{D}$-odd components.
All the other components are $P'_\mathrm{D}$-even.

Therefore, the gauge bosons $A^2$ and $A^3$ together with their corresponding Goldstone bosons $G_2$ and $G_3$ are $P'_\mathrm{D}$-odd, since $A^2$ and $A^3$ ($G_2$ and $G_3$) are linear combinations of $\tilde{A}^2$ and $\tilde{A}^3$ ($\phi_{1}$ and $\chi_1$).
The rest physical states are all $P'_\mathrm{D}$-even.
Consequently, the lightest gauge boson $A^2$ cannot decay, serving as a DM candidate.
Generally when $m_{A^3} \neq m_{A^2}$, $A^3$ finally decays into $A^2$,  however it could coannihilate with $A^2$ during the freeze-out epoch if $m_{A^3} \sim m_{A^2}$.

\section{Phenomenology}
\label{sec:pheno}

In this section, we scan the parameter space, taking into account the constraints from the observed DM relic density and the collider measurements of the 125~GeV Higgs boson.
DM scattering and annihilation cross sections for direct and indirect detection predicted by this model is further calculated.
We describe the details of the parameter scan and discuss the numerical results in the following subsections.

\subsection{Parameter scan and DM relic density}

We adopt the following 14 real parameters,
\begin{equation}
  \label{eq: Free parameters}
  \gd ,\  \lambda_0 ,\  \lambda_1 ,\  \lambda_2 ,\  \lambda_3 ,\  \lambda_4 ,\  \lambda_5 ,\  \lambda_6 ,\  \lambda_{10} ,\  \lambda_{20} ,\  \lambda_{30} ,\  v_1 ,\  v_2 ,\  v_3,
\end{equation}
as the independent parameters. A random scan is carried out for the parameters in logarithmic scales within the following range:
\begin{eqnarray}
    &&10^{-3} < \gd < 1 , \qquad
    10~\si{GeV} < v_1 ,\  v_2 ,\  v_3 < 10^3~\si{GeV},
\\
    &&10^{-3} < \lambda_0 ,\ \lambda_1 ,\ \lambda_2,\ |\lambda_3| ,\ |\lambda_4| ,\ |\lambda_5| ,\ |\lambda_6| ,\ |\lambda_{10}| ,\ |\lambda_{20}| ,\ |\lambda_{30}| < 1 .
\end{eqnarray}
Note that the positivity of the $m_{h_i}^2$ demands the positive $\lambda_0$, $\lambda_1$, and $\lambda_2$.
We numerically calculate the eigenvalues and eigenvectors of $\mathcal{M}_h^2$, and construct $\mathcal{O}_h$.
We require that the SM-like Higgs boson $h_0$ has the most contribution from $\tilde{h}_0$, and that $m_{h_0}$ lies within the $3 \sigma$ range of the measured value $125.10 \pm 0.14~\si{GeV}$~\cite{Zyla:2020zbs}.
Other eigenvalues of $\mathcal{M}_h^2$ are sorted in an ascending order and accordingly appointed to $h_1$, $h_2$, and $h_3$.

We implement the Lagrangian in \texttt{FeynRules~2.3.36}~\cite{Alloul:2013bka} to generate the \texttt{CalcHEP} input files, and feed them to \texttt{microOMEGAs~5.4}~\cite{Belanger:2006is, Belanger:2008sj, Barducci:2016pcb} for phenomenological calculations.
Based on current Higgs measurements at the Large Hadron Collider (LHC), the SM-like Higgs boson is tested by \texttt{Lilith~2.0}~\cite{Bernon:2015hsa, Kraml:2019sis} implemented in the \texttt{micrOMEGAs}, and the corresponding $p$-values are derived.
Parameter points with $p$-values less than $0.05$ are discarded, corresponding to the exclusion at 95\% confidence level (C.L.).

The exotic neutral Higgs bosons $h_1$, $h_2$, and $h_3$ in this model might be directly produced at the LEP and the LHC.
Thus, the parameter points are also constrained by the direct searches at these colliders, which are utilized to constrain the mixing angle $\sin\theta$ of a second neutral Higgs boson in Fig.~3 of Ref.~\cite{Falkowski:2015iwa}.
As a good approximation, we reinterpret these constraints on $|\sin \theta|$ as the constraints on $|\mathcal{O}_{h,i0}| \sqrt{\mathrm{Br} (h_i \rightarrow \mathrm{SM})}$ ($i = 1,2,3$), where $\mathrm{Br} (h_i \rightarrow \mathrm{SM})$ is the $h_i$ decay branching ratio to the SM final states including $h_0$, and $\sqrt{\mathrm{Br} (h_i \rightarrow \mathrm{SM})}$ acts as a rescaling factor of the signal strengths for comparison with the collider data.

In addition, these exotic Higgs bosons couple to the $W$ and $Z$ bosons through their components of the $\mathrm{SU(2)_L}$ Higgs doublet, inducing one-loop corrections to the propagators of the electroweak gauge bosons.
The corresponding shifts of the $g^{\mu\nu}$ coefficients of the vacuum polarization amplitudes with respect to the SM are given by
\begin{eqnarray}
\delta \Pi_{WW}(p^2) &=& \frac{m_W^2}{4 \pi^2 v_0^2} \bigg\{ \sum_{i=0}^3 \mathcal{O}_{h,i0}^2 \left[\frac{m_{h_i}^2}{4}\ln m_{h_i}^2 + F(p^2, m_W^2, m_{h_i}^2) \right]
 - \frac{m_{h_\mathrm{SM}}^2}{4}\ln m_{h_\mathrm{SM}}^2
\nonumber\\
&&\qquad\qquad\quad~ - F(p^2, m_W^2, m_{h_{\mathrm{SM}}}^2) \bigg\},
\\
\delta \Pi_{ZZ}(p^2) &=& \frac{m_Z^2}{4 \pi^2 v_0^2} \bigg\{ \sum_{i=0}^3 \mathcal{O}_{h,i0}^2 \left[\frac{m_{h_i}^2}{4}\ln m_{h_i}^2 + F(p^2, m_Z^2, m_{h_i}^2) \right]
- \frac{m_{h_\mathrm{SM}}^2}{4}\ln m_{h_\mathrm{SM}}^2
\nonumber\\
&&\qquad\qquad\quad~  - F(p^2, m_Z^2, m_{h_{\mathrm{SM}}}^2) \bigg\},
\\
\delta \Pi_{\gamma \gamma}(p^2) &=& \delta \Pi_{Z\gamma}(p^2) =0,
\end{eqnarray}
where $m_{h_\mathrm{SM}}$ is the SM Higgs boson mass  and the loop function $F$ is defined by~\cite{Falkowski:2015iwa}
\begin{eqnarray}
F(p^2, m_1^2, m_2^2) = \int_0^1 dx \left(m_1^2-\frac{\Delta}{2}\right) \ln \Delta,
\end{eqnarray}
with
\begin{eqnarray}
\Delta = x m_2^2 + (1-x) m_1^2 - p^2 x(1-x).
\end{eqnarray}
Following the discussions in Ref.~\cite{Falkowski:2015iwa}, we calculate the shifts of the precisely measured electroweak quantities $\Gamma_Z$, $R_{\ell}$, $R_{b}$, $\sin^2 \theta_{\mathrm{eff}}^\ell$, and $m_W$, and compare the shifted values with the $2\sigma$ ranges of the experimental values adopted from Ref.~\cite{Zyla:2020zbs} to filter all the parameter points.

Now we discuss the DM relic density predicted in this model, where the DM particles are $A^2$ gauge bosons.
Since the $P'_\mathrm{D}$-odd $A^3$ must decay into $A^2$, its abundance can also contribute to the DM relic density.
If $m_{A^3} - m_{A^2} \lesssim 0.1 m_{A^2}$, the $A^{2,3}$ coannihilation processes at the freeze-out epoch could significantly affect the relic density~\cite{Griest:1990kh}.
Therefore, we consider all possible annihilation and coannihilation channels with two-body final states, as listed below.
\begin{enumerate}
  \item $A^{2,3} A^{2,3} \to h_i h_j$, induced by  $s$-channel Higgs bosons and quartic couplings.
  \item $A^{2,3} A^{2,3} \to f\bar{f}$, mediated by $s$-channel Higgs bosons. $f$ denotes SM fermions.
  \item $A^{2,3} A^{2,3} \to W^+ W^-, Z^0 Z^0$, mediated by $s$-channel Higgs bosons.
\end{enumerate}
We utilize \texttt{micrOMEGAs} to evaluate the relic density, including the coannihilation effect.

\begin{figure}
  \centering
  \includegraphics[width=0.495\textwidth]{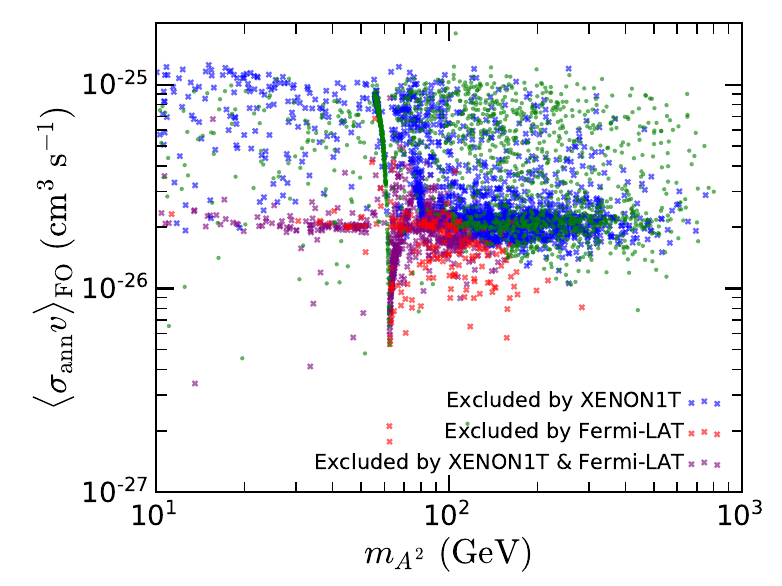}
\caption{Parameter points predicting the observed DM relic density projected in the $m_{A^2}$-$\sv_{\mathrm{FO}}$ plane.
The blue and red points are excluded by the XENON1T direct detection experiment and the Fermi-LAT indirect detection experiment, respectively, while the purple points are excluded by both.
The green points survive from both constraints.
The details will be described in subsection~\ref{subsec:DD&ID}.}
\label{fig: mA2 svAnnFO svAnn}
\end{figure}

In \texttt{micrOMEGAs}, the freeze-out temperature $T_\mathrm{f}$ indicating the departure from thermal equilibrium is defined by $Y(T_\mathrm{f}) = 2.5 Y_{\mathrm{eq}}(T_\mathrm{f})$, where $Y = n_\mathrm{DM}/s$ is the DM number density $n_\mathrm{DM}$ divided by the entropy density $s$, and $Y_\mathrm{eq}$ is its equilibrium value.
We derive the thermally averaged effective annihilation cross section times velocity $\sv_{\mathrm{FO}}$ at $T = T_\mathrm{f}$ for the parameter points, as shown in Fig.~\ref{fig: mA2 svAnnFO svAnn}.
All the points in the figure predict the relic density $\Omega_{\mathrm{DM}} h^2$ within the $3\sigma$ range of the Planck measured value $0.1200 \pm 0.0012$~\cite{Aghanim:2018eyx}.

Besides the most prominent points with the basically stable $\sv$ so that $\sv_{\mathrm{FO}}$ is close to the  standard annihilation cross section $\sv_\mathrm{sd} \simeq 2 \times 10^{-26}~\si{cm^3~s^{-1}}$,
there are enormous points with significantly temperature-dependent $\sv$, so their $\sv_{\mathrm{FO}}$ deviate from the standard value. Part of the deviations is due to the Breit-Wigner resonance effects~\cite{Griest:1990kh}.

In Fig.~\ref{fig: mA2 svAnnFO svAnn}, there is a significant resonant structure lying around $m_{A^2} \sim m_{h_0} / 2 \simeq  62.5~\si{GeV}$ due to the $A^2 A^2$ annihilation mediated by the $s$-channel SM-like Higgs boson $h_0$. For the points with $\sv_{\mathrm{FO}} < \sv_\mathrm{sd}$, the invariant mass $m_\mathrm{inv}$ of the $A^2 A^2$ pairs is typically higher than the resonance center $m_{h_0}$ at the $T_\mathrm{f}$, and then decreases to approach the $m_{h_0}$ as the temperature drops, lifting the $\sv$ to consume more DM particles in the later period. Therefore, a smaller $\sv_{\mathrm{FO}}$ is needed for the observed relic density. On the other hand, some other points with $\sv_{\mathrm{FO}} > \sv_\mathrm{sd}$ probably have the $m_\mathrm{inv}$ equal to or slightly lower than the $m_{h_0}$ at the $T_\mathrm{f}$, Lowering the temperature will therefore increase the distance between $m_{h_0}$ and the invariant mass of the annihilating $A^2 A^2$ pairs, suppressing the annihilation at lower temperatures. Hence, the observed relic density requires a larger $\sv_{\mathrm{FO}}$.

\begin{figure}
  \centering
  \subfigure[~$\sv_{\mathrm{FO}}$ versus $m_{A^2}/m_{h_1}$.\label{fig:mA2-mHi_svAnnFO:h1}]{\includegraphics[width=0.495\textwidth]{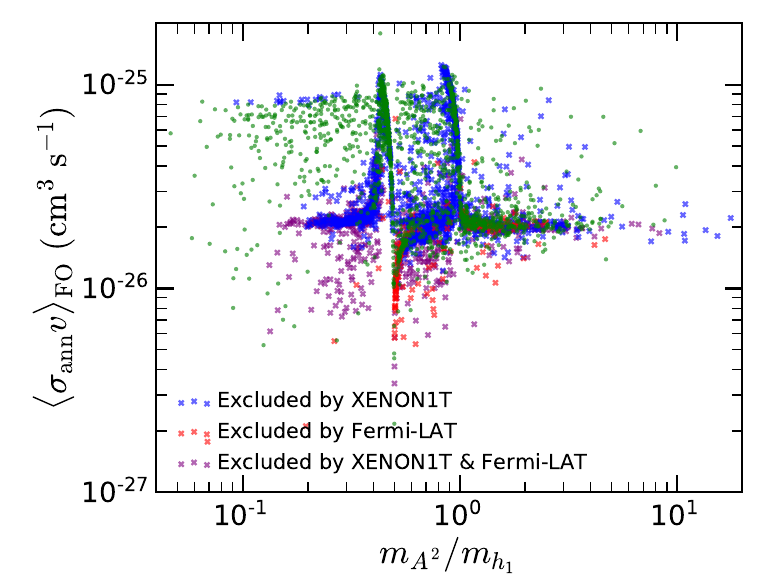}}
  \subfigure[~$\sv_{\mathrm{FO}}$ versus $m_{A^2}/m_{h_2}$.\label{fig:mA2-mHi_svAnnFO:h2}]{\includegraphics[width=0.495\textwidth]{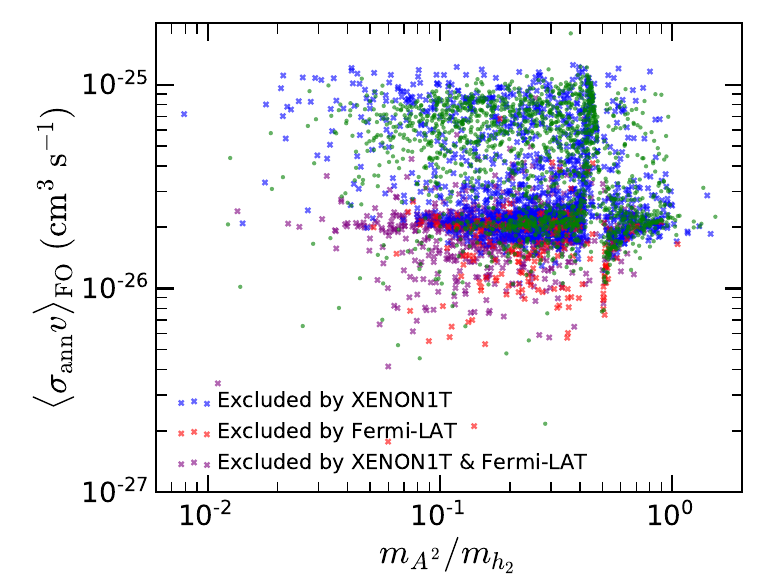}}
  \subfigure[~$\sv_{\mathrm{FO}}$ versus $m_{A^2}/m_{h_3}$.\label{fig:mA2-mHi_svAnnFO:h3}]{\includegraphics[width=0.495\textwidth]{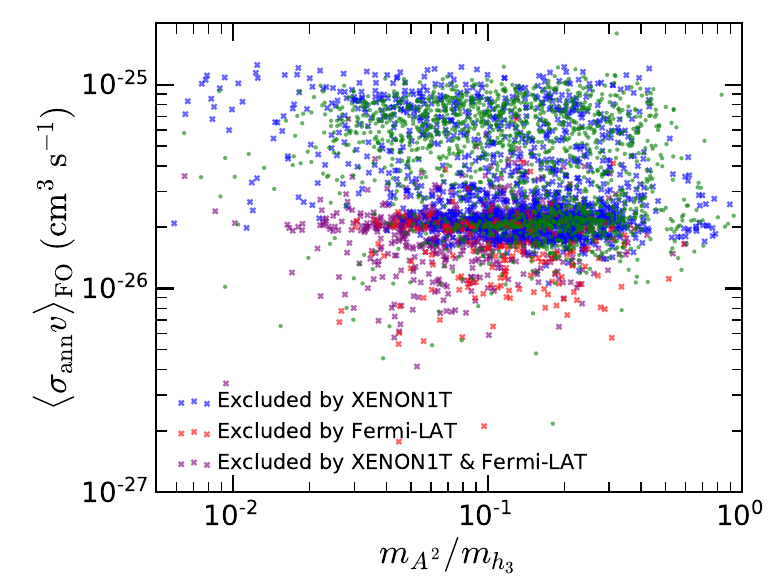}}
\caption{Freeze-out effective annihilation cross section $\sv_{\mathrm{FO}}$ versus $m_{A^2}/m_{h_1}$ (a), $m_{A^2}/m_{h_2}$ (b), and $m_{A^2}/m_{h_3}$ (c).}
\label{fig:mA2-mHi_svAnnFO}
\end{figure}

Besides the SM-like Higgs boson, the exotic Higgs bosons $h_1$, $h_2$, and $h_3$ can also give rise to resonance effects.
To see this clearly, we change the horizontal axis to $m_{A^2}/m_{h_{1,2,3}}$ and plot the points in Fig.~\ref{fig:mA2-mHi_svAnnFO}.
Remarkable structures due to resonance effects appear at $m_{A^2}/m_{h_1} \sim 1/2$ in Fig.~\ref{fig:mA2-mHi_svAnnFO:h1} and at $m_{A^2}/m_{h_2} \sim 1/2$ in Fig.~\ref{fig:mA2-mHi_svAnnFO:h2}.
Nevertheless, Fig.~\ref{fig:mA2-mHi_svAnnFO:h3} does not show an obvious resonance structure,
because $h_3$ is the heaviest exotic Higgs boson, and $m_{A^2}$ is probably too small to approach $m_{h_3}/2$ in our scan.

In Fig.~\ref{fig:mA2-mHi_svAnnFO:h1}, there is a ``peak'' structure near ${m_{A^2}}/m_{h_{1}} \sim 1$.
This originates from the $h_1 h_1$ threshold effect~\cite{Griest:1990kh}.
If $m_{A^2}$ is slightly smaller than $m_{h_1}$, the $A^2 A^2 \rightarrow h_1 h_1$ annihilation channel would open at sufficiently high temperatures,  and then would be kinematically prohibited as the temperature drops, leading to a rapid shrink of $\sv$ to cease further annihilation, requiring a larger $\sv_{\mathrm{FO}}$ to clear out the redundant dark matter in advance.

\begin{figure}
  \centering
  \includegraphics[width=0.495\textwidth]{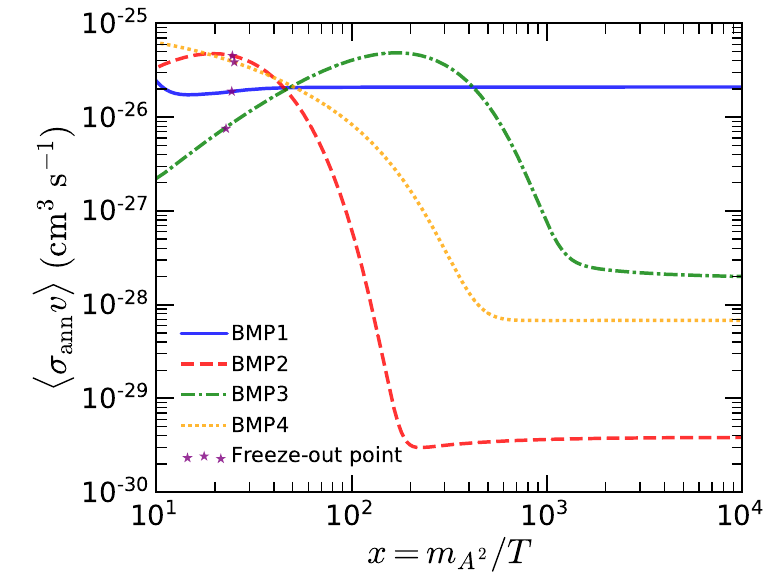}
\caption{Effective annihilation cross sections $\sv$ as functions of $x= m_{A^2}/T$ for four benchmark points.
The freeze-out point for each curve is indicated by a star mark.}
  \label{fig: threshold}
\end{figure}

To see both the resonance and the threshold effects clearly, we select four benchmark points (BMPs) and plot their $\sv$ as functions of $x\equiv m_{A^2}/T$ in Fig.~\ref{fig: threshold}.
The properties of the BMPs are listed in Table~\ref{tab:benchmark-point}.
BMP1 is a normal benchmark point without these effects, leading to roughly constant $\sv$ from $x = 10$ to $x = 10^4$.
For BMP2 with $m_{A^2} \sim m_{h_1}/2$, the $h_1$ resonance effect is important.
As a result, $\sv$ is larger than
$\sv_\mathrm{sd}$ at the freeze-out epoch, but suddenly decline since then.
BMP3 with $m_{A^2} \sim m_{h_0}/2$ is affected by the $h_0$ resonance, leading to $\sv_\mathrm{FO} < \sv_\mathrm{sd}$.
After $x_\mathrm{f} = m_{A^2}/T_\mathrm{f}$, its $\sv$ is elevated by the more effective resonance, and then peaks at $x \sim 200$, and finally decreases because the invariant masses of the $A^2$ pairs drop to pass through the $m_{h_0}$ value to reduce the resonant effect again.
The $h_1h_1$ threshold effect manifests for BMP4 with $m_{A^2} \sim m_{h_1}$, and hence $\sv$ decreases as the temperature decreases until $x \sim 400$.

\begin{table}
  \centering
  \def\arraystretch{1.2}
\caption{Properties of four benchmark points, corresponding to four curves in Fig.~\ref{fig: threshold}.}
\label{tab:benchmark-point}
  \begin{tabular}{|C{0.22\textwidth}C{0.17\textwidth}C{0.18\textwidth}C{0.17\textwidth}C{0.17\textwidth}|}
\hline
                                                   &               BMP1 &               BMP2 &               BMP3 &               BMP4 \\
\hline
$g_\mathrm{D}$                                     & \num{       0.232} & \num{       0.392} & \num{        0.190} & \num{       0.293} \\
$\lambda_0$                                        & \num{        0.130} & \num{       0.171} & \num{       0.129} & \num{       0.128} \\
$\lambda_1$                                        & \num{       0.112} & \num{       0.757} & \num{      0.0134} & \num{       0.431} \\
$\lambda_2$                                        & \num{      0.0631} & \num{       0.0830} & \num{      0.0312} & \num{      0.0103} \\
$\lambda_3$                                        & \num{     0.00144} & \num{     -0.00810} & \num{     0.00362} & \num{     0.00877} \\
$\lambda_4$                                        & \num{     0.00654} & \num{     -0.0367} & \num{     -0.0228} & \num{     -0.0616} \\
$\lambda_5$                                        & \num{     0.00795} & \num{     -0.0207} & \num{       -0.0200} & \num{     0.00587} \\
$\lambda_6$                                        & \num{     0.00177} & \num{      0.0414} & \num{       0.136} & \num{       0.578} \\
$\lambda_{10}$                                     & \num{      0.0124} & \num{      0.0353} & \num{    -0.00189} & \num{     -0.0574} \\
$\lambda_{20}$                                     & \num{     0.00105} & \num{      -0.108} & \num{    -0.00107} & \num{      0.0024} \\
$\lambda_{30}$                                     & \num{     0.00117} & \num{     0.00371} & \num{     -0.0115} & \num{     0.00621} \\
$v_1~\si{(GeV)}$                                   & \num{         714} & \num{         179} & \num{         692} & \num{         973} \\
$v_2~\si{(GeV)}$                                   & \num{         647} & \num{         485} & \num{         353} & \num{         410} \\
$v_3~\si{(GeV)}$                                   & \num{        35.3} & \num{          12.0} & \num{         247} & \num{         204} \\
$m_{A^1}~\si{(GeV)}$                               & \num{         224} & \num{         203} & \num{         155} & \num{         315} \\
$m_{A^2}~\si{(GeV)}$                               & \num{         149} & \num{        70.2} & \num{        62.2} & \num{         117} \\
$m_{A^3}~\si{(GeV)}$                               & \num{         167} & \num{         190} & \num{         142} & \num{         293} \\
$m_{h_1}~\si{(GeV)}$                               & \num{        51.3} & \num{         147} & \num{         182} & \num{         118} \\
$m_{h_2}~\si{(GeV)}$                               & \num{         462} & \num{         402} & \num{         215} & \num{    1.14e+03} \\
$m_{h_3}~\si{(GeV)}$                               & \num{         676} & \num{         441} & \num{         412} & \num{    1.81e+03} \\
$x_\mathrm{f}\equiv m_{A^2}/T_\mathrm{f}$          & \num{        24.4} & \num{        24.5} & \num{        22.8} & \num{        25.1} \\
$\sv_\mathrm{FO}~\si{(cm^3/s)}$               & \num{    1.88e-26} & \num{    4.52e-26} & \num{    7.55e-27} & \num{    3.87e-26} \\
$\sv_0~\si{(cm^3/s)}$                         & \num{     2.10e-26} & \num{    3.89e-30} & \num{    1.93e-28} & \num{    6.96e-29} \\
$\sigma_N^\mathrm{SI}~\si{(cm^2)}$                 & \num{    2.02e-47} & \num{    1.41e-47} & \num{    1.04e-50} & \num{    8.58e-47} \\
$\Omega_\mathrm{DM} h^2$                           & \num{       0.122} & \num{       0.118} & \num{       0.117} & \num{       0.117} \\

\hline
\end{tabular}
\end{table}

\subsection{Direct and indirect detection}
\label{subsec:DD&ID}

In this subsection, we calculate the predictions for direct and indirect detection and compare them with the experimental data.

In the model, the DM particle $A^2$ interacts with quarks ($q = d, u, s, c, b, t$) via interchanging all the Higgs bosons.
The relevant Higgs-portal interaction terms can be expressed as
\begin{equation}
  \Lag_\mathrm{portal} = \sum_{i = 0}^3 \left(\frac{\kappa_i v_0}{2}\, h_i A^2_{\mu} A^{2,\mu} - \sum_{q} \frac{\mathcal{O}_{h,0i} m_q}{v_0}\, h_i \bar{q} q\right),
\end{equation}
where
\begin{equation}
     \kappa_i = \frac{2 \gd^2}{v_0} \left\{ \frac{\mathcal{O}_{h,1i}}{v_{23}} (s_{\theta} v_2 - c_{\theta} v_3)^2 - \frac{\mathcal{O}_{h,2i} v_1 v_2 }{v_{23} v_{123}}  (s_{2\theta} v_2 - c_{2\theta} v_3) + \mathcal{O}_{h,3i} c_{\theta}^2 v_1  \right\}.
\end{equation}
The $A^2$-quark interactions induce $A^2$-nucleon interactions.
Thus, the spin-independent (SI) cross section for $A^2$ scattering off a nucleon $N = p, n$ is given by~\cite{Yu:2011by}
\begin{equation}
  \sigma_{A^2 N} = \frac{G_{A^2 N}^2 \mu_{A^2, N}^2}{4 \pi m_{A^2}^2} ,
\end{equation}
with the $A^2$-$N$ reduced mass defined as $\mu_{A^2, N} = m_{A^2} m_N / (m_{A^2} + m_N)$ and the effective coupling
\begin{equation}
  G_{A^2 N} = - m_N \sum_{q} f_q^N \sum_{i=0}^{3}  \frac{\kappa_i \mathcal{O}_{h,0i}}{m_{h_i}^2}.
\end{equation}
$f_q^N$ are the nucleon form factors for quarks~\cite{Belanger:2008sj}.

In this model, $\sigma_{A^2 n}$ is slightly different from $\sigma_{A^2 p}$.
In order to take this difference into account, we should calculate the normalized-to-nucleon SI cross section~\cite{Feng:2011vu}
\begin{equation}\label{eq:nor-SI-cross-section}
  \sigma_N^\mathrm{SI} = \sigma_{A^2 p}\, \dfrac{ \sum\limits_I \eta_I \mu_{A^2,A_I}^2 \left[ Z + (A_I - Z) {G_{A^2n}}/{G_{A^2p}} \right]^2 }{ \sum\limits_I \eta_I \mu_{A^2,A_I}^2 A_I^2 }
\end{equation}
for $A^2$ scattering off nuclei $A_I$ with $Z$ protons and $A_I - Z$ neutrons.
The index $I$ denotes isotopes, and $\eta_I$ is the fractional number abundance of the isotope $A_I$ in nature.

\begin{figure}
  \centering
  \subfigure[~$m_{A^2}$-$\sigma_N^\mathrm{SI}$ plane.\label{fig: Exclusion plots' XENON1T}]{\includegraphics[width=0.495\textwidth]{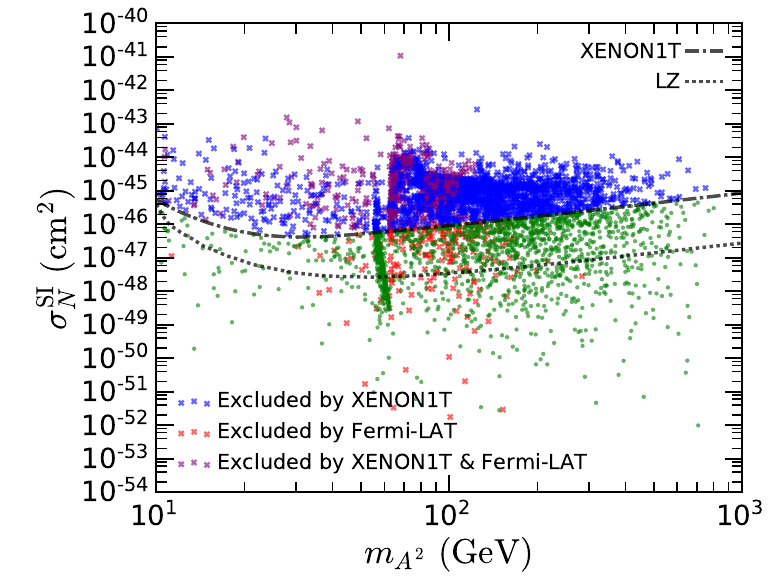}}
  \subfigure[~$m_{A^2}$-$\sv_0$ plane.\label{fig: Exclusion plots' Fermi-LAT}]{\includegraphics[width=0.495\textwidth]{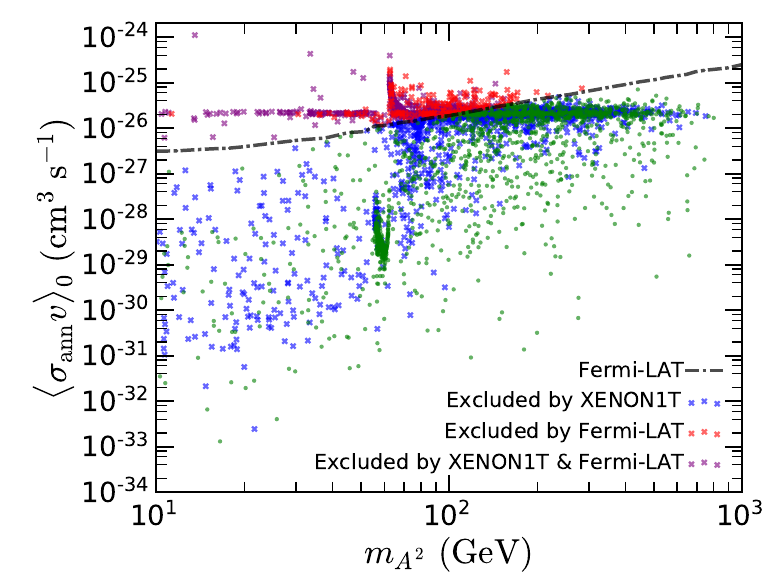}}
\caption{Parameter points projected in the $m_{A^2}$-$\sigma_N^\mathrm{SI}$ (a) and $m_{A^2}$-$\sv_0$ (b) planes.
In the left panel, the dot-dashed line denotes the $90\%$ C.L. upper limit on $\sigma_N^\mathrm{SI}$ from XENON1T~\cite{Aprile:2018dbl}, while the dotted line demonstrates the $90\%$ C.L. projected sensitivity of LZ~\cite{Mount:2017qzi}.
In the right panel, the dot-dashed line indicates the $95\%$ C.L. upper limit from the Fermi-LAT $\gamma$-ray observations of dwarf galaxies~\cite{Hoof:2018hyn}.
The blue points are excluded by XENON1T, the red points excluded by Fermi-LAT, the purple points are excluded by both, and the green points survive from the two bounds.}
\label{fig: Exclusion plots}
\end{figure}

We can compare $\sigma_N^\mathrm{SI}$ predicted by the model with the 90\% C.L. upper limits on $\sigma_N^\mathrm{SI}$ from the direct detection experiment XENON1T~\cite{Aprile:2018dbl}.
The detection material in XENON1T is xenon ($Z=54$), whose main isotopes have $A_I = \{128$, $129$, $130$, $131$, $132$, $134$, $136\}$ with $\eta_I = \{1.9\%$, $26\%$, $4.1\%$, $21\%$, $27\%$, $10\%$, $8.9\%\}$~\cite{Feng:2011vu}.
We compute $\sigma_N^\mathrm{SI}$ for the parameter points and plot the result in Fig.~\ref{fig: Exclusion plots' XENON1T}. The dot-dashed line indicates the XENON1T bound, which excludes a fraction of the parameter points.
This model can also be tested in future direct detection experiments.
For an example, the proposed LZ detector is designed to reach a sensitivity of $\sim 3 \times 10^{-48}\ \mathrm{cm^2}$ for $\sigma^{\mathrm{SI}}_N$~\cite{Mount:2017qzi}, which is expected to be relevant to some parameter regions of the model.
We also plot the LZ sensitivity as a dotted line in Fig.~\ref{fig: Exclusion plots' XENON1T}.

Now we discuss indirect detection.
$A^2 A^2$ annihilation can produce high energy $\gamma$ rays, which might be received by the Fermi Large Area Telescope (Fermi-LAT).
It has long been known that the dwarf spheroidal galaxies surrounding the Milky Way Galaxy are dominated by dark matter, as their mass-to-light ratios are of a hundred or greater~\cite{Strigari:2013iaa}.
A global analysis from $\gamma$-ray observations of 27 dwarf galaxies using 11 years of the Fermi-LAT data~\cite{Hoof:2018hyn} offers a stringent bound on DM annihilation.

In Fig.~\ref{fig: mA2 svAnnFO svAnn}, we project the parameter points in the $m_{A^2}$-$\sv_0$ plane, where $\sv_0$ is the $A^2 A^2$ annihilation cross section in the low velocity limit.
The dot-dashed line denotes the $95\%$ C.L. upper limit on the DM annihilation cross section in the $b\bar{b}$ channel derived by the analysis of Fermi-LAT data.
Since the annihilation channels into $W^+ W^-$, $Z Z$, $t\bar{t}$, and a pair of Higgs bosons induce  $\gamma$-ray spectra similar to that from the $b\bar{b}$ channel~\cite{Cirelli:2010xx}, we can approximately compare $\sv_0$ with the Fermi-LAT $b \bar{b}$ constraint.

In Figs.~\ref{fig: mA2 svAnnFO svAnn}, \ref{fig: threshold}, and \ref{fig: Exclusion plots}, we have denoted the parameter points with colors.
The blue and red points are excluded by XENON1T and Fermi-LAT, respectively, while the purple points are excluded by both.
The green points survive from the two constraints.
Comparing Fig.~\ref{fig: Exclusion plots' Fermi-LAT} with Fig.~\ref{fig: mA2 svAnnFO svAnn}, one may notice that $\sv_0 \ll \sv_{\mathrm{FO}}$ holds for many points.

As mentioned above, when the temperature drops, the  $A^2 A^2$ invariant mass decreases and the distance to a resonance center changes.
Consequently, for the parameter points with $m_{A^2} \sim {m_{h_0}}/{2}$, which are related to the $h_0$ resonance, $\sv$ at low temperatures significantly decreases for $m_{A^2} < {m_{h_0}}/{2}$, but could remarkably increase for $m_{A^2} > {m_{h_0}}/{2}$.
This is clearly shown in Fig.~\ref{fig: Exclusion plots' Fermi-LAT}, where the parameter points in the latter case lead to too large $\sv_0$,  usually excluded by the Fermi-LAT.
For the former case, both $\sigma_N^\mathrm{SI}$ and $\sv_0$ could be small enough to evade the XENON1T and Fermi-LAT constraints, as demonstrated in Figs.~\ref{fig: Exclusion plots' XENON1T} and \ref{fig: Exclusion plots' Fermi-LAT}.

\begin{figure}
\includegraphics[width=0.495\textwidth]{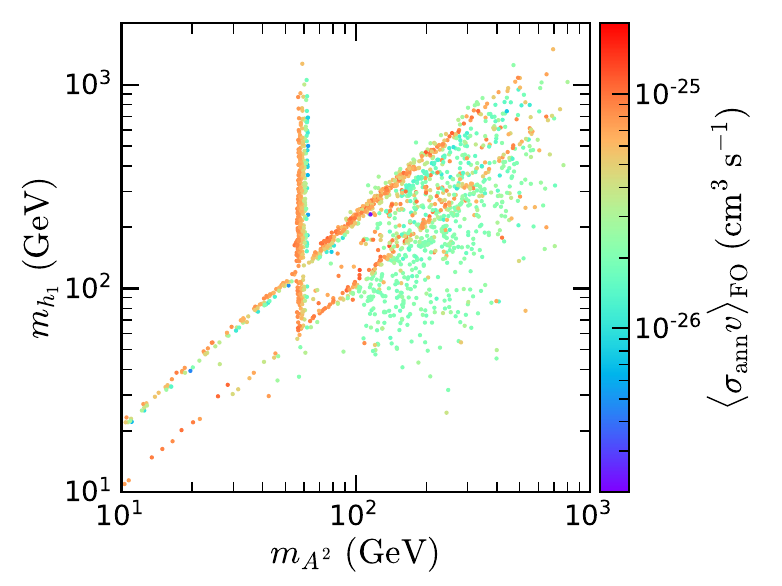}
\includegraphics[width=0.495\textwidth]{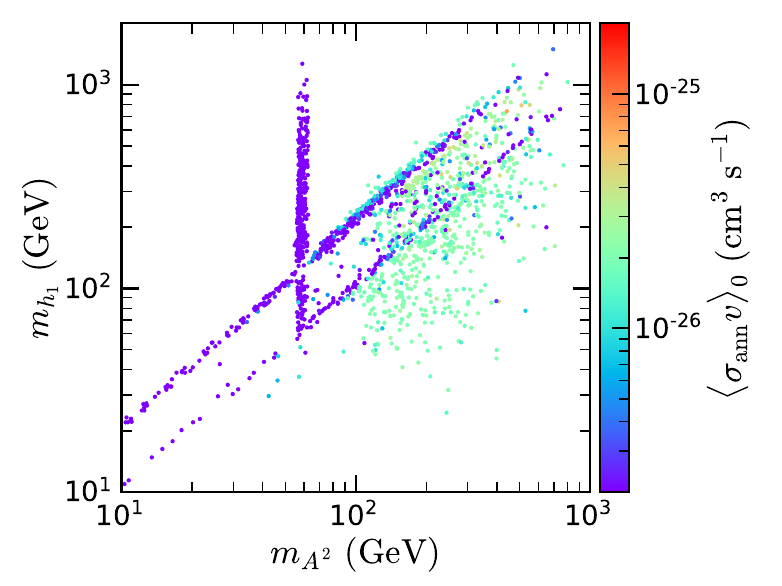}
\includegraphics[width=0.495\textwidth]{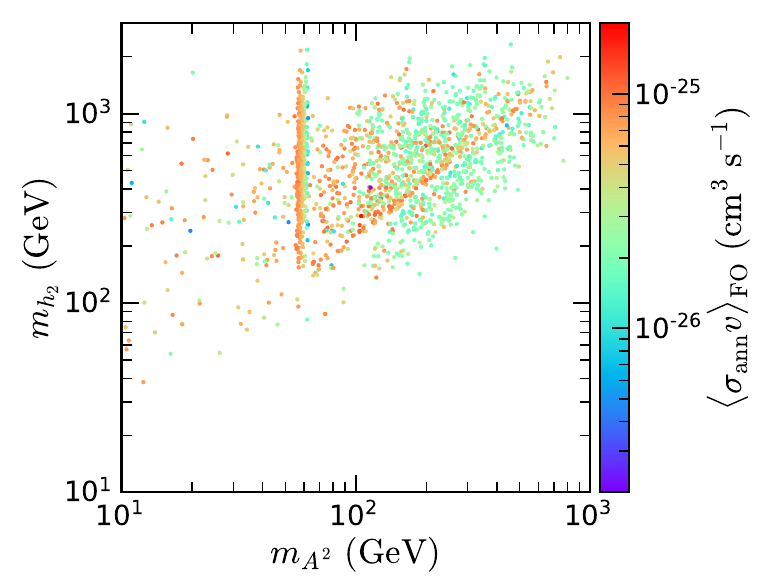}
\includegraphics[width=0.495\textwidth]{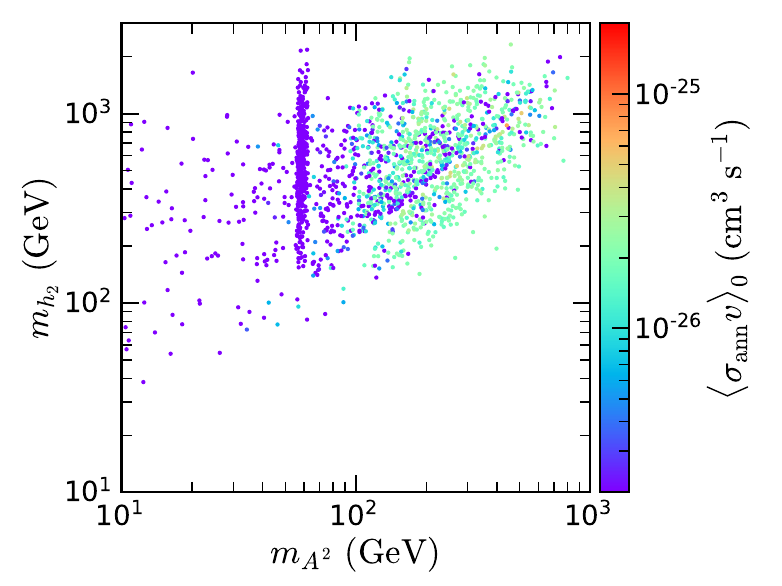}
\caption{$\sv_{\mathrm{FO}}$ (left panels) and $\sv_0$ (right panels) for the surviving parameter points projected in the $m_{A^2}$-$m_{h_1}$ (upper panels) and $m_{A^2}$-$m_{h_2}$ (lower panels) planes.}
\label{FO_No_Compare}
\end{figure}

The resonances of the exotic Higgs bosons have similar effects.
To illustrate it, we compare $\sv_{\mathrm{FO}}$ and $\sv_0$ as the color axes in Fig.~\ref{FO_No_Compare}, where only the parameter points survive from both the XENON1T and Fermi-LAT constraints are shown in the $m_{A^2}$-$m_{h_1}$ and $m_{A^2}$-$m_{h_2}$ planes.
In the $m_{A^2}$-$m_{h_1}$ plane, the points roughly align along three lines.
The first line with $m_{A^2} \sim 62.5~\si{GeV}$ corresponds to the $h_0$ resonance effect, as we just explained.
The second line with $m_{A^2} \sim m_{h_1}/2$ corresponds to the $h_1$ resonance effect for a similar reason.
The third line with $m_{A^2} \sim m_{h_1}$ is related to the $h_1 h_1$ threshold effect, as discussed for Fig.~\ref{fig:mA2-mHi_svAnnFO:h1} in the previous subsection.
Because of these effects, $\sv_0$ is much smaller than $\sv_{\mathrm{FO}}$ along these lines.
In the $m_{A^2}$-$m_{h_2}$ plane, likewise, there is a line with $m_{A^2} \sim m_{h_2}/2$ showing a similar behavior.
Nonetheless, the line with $m_{A^2} \sim m_{h_2}$ does not manifest, because we have just a few points in this region.

\section{Generalization to the general $\mathrm{SO}(N)$ cases}
\label{sec:extension}

In this model, we introduce a dark $\mathrm{SU(2)_D} \simeq \mathrm{SO(3)_D}$ gauge symmetry broken by two real $\mathrm{SU(2)_D}$ Higgs triplets.
As a result, the accidental $Z_2$ symmetry extends the global symmetry of the Lagrangian to $\mathrm{O(3)_D}$.
After the spontaneous symmetry breaking,  a $Z'_2$ symmetry remains, leading to two $P'_\mathrm{D}$-odd dark gauge bosons, where the lighter one serves as a vector DM candidate.
Such a setup can be generalized to a dark $\mathrm{SO}(N)_\mathrm{D}$ gauge group with $N > 3$.
We prove the possibility in this section.

A $\mathrm{SO}(N)_\mathrm{D}$ gauge symmetry can be completely and exactly broken by $N-1$ real Higgs multiplets in the $N$-dimensional fundamental representation. Similarly, if the Lagrangian respects a global $\mathrm{O}(N)_\mathrm{D}$ symmetry accidentally, after the spontaneous symmetry breaking, all the $N-1$ linearly independent VEVs of these Higgs multiplets can determine a $(N-1)$-dimensional hypersurface in the representation space, and the reflection operation with respect this hypersurface indicates a $Z'_2$ symmetry remaining to guarantee the stability of the dark gauge boson. The validation of the $\mathrm{O}(N)_\mathrm{D}$ symmetry of the Lagrangian requires a dark parity $P_\mathrm{D}$ matrix with $\det(P_\mathrm{D}) = -1$, and $P_{\mathrm{D}} P_{\mathrm{D}}^{\dagger} = I$. If one can prove that all the possible renormalizable $\mathrm{SO}(N)_\mathrm{D}$-invariant terms remain $P_{\mathrm{D}}$-even, then the $\mathrm{O}(N)_\mathrm{D}$ symmetry is respected accidentally.

If there are $N$ Higgs multiplets $\Phi_i$ as $\mathrm{O}(N)_\mathrm{D}$ vectors instead,  one can construct a pseudoscalar term $\varepsilon^{a_1 a_2 \cdots a_N} \Phi_1^{a_1} \Phi_2^{a_2} \cdots \Phi_N^{a_N}$ with the $N$-dimensional Levi-Civita symbol. Notice that
\begin{equation}
\varepsilon^{a_1 a_2 \cdots a_N} = \det (R) R_{a_1 b_1} R_{a_2 b_2} \cdots R_{a_N b_N} \varepsilon^{b_1 b_2 \cdots b_N}
\end{equation}
for $R\in \mathrm{O}(N)_\mathrm{D}$, so such a term is $P_\mathrm{D}$-odd.
Therefore,  only the global $\mathrm{SO}(N)_\mathrm{D}$ symmetry is respected for the $N$ Higgs multiplets.
However,  we will prove by contradiction that no $P_\mathrm{D}$-odd potential term arises for the $N-1$ Higgs multiplets.  This can be done by verifying that $\mathrm{O}(N)_\mathrm{D}$ vectors $V_1, V_2, \cdots, V_n$ with $n<N$ cannot form any $P_\mathrm{D}$-odd scalar.

Assume that $V_1, V_2, \cdots, V_n$ ($n<N$) can form a $P_\mathrm{D}$-odd scalar.
Then, similarly as the Levi-Civita symbol, there must exist a ``constant $\mathrm{O}(N)_\mathrm{D}$ tensor'' $E^{a_1 a_2 \cdots a_n}$ leading to a $P_\mathrm{D}$-odd scalar $E^{a_1 a_2 \cdots a_n} V_1^{a_1} V_2^{a_2} \cdots V_n^{a_n}$, where $a_i$ ($i=1,2,\cdots,n$) are $\mathrm{O}(N)_\mathrm{D}$ indices.
The tensor components $E^{a_1 a_2 \cdots a_n}$ are constants because they are the combinations of the corresponding Clebsch–Gordan coefficients which stay constantly as the coordinates rotate.  That is to say, the tensor $E^{a_1 a_2 \cdots a_n}$ is {\it invariant} under any $\mathrm{SO}(N)_\mathrm{D}$ transformation, so we have
\begin{eqnarray}
E^{a_1 a_2 \cdots a_n} = R_{a_1 b_1} R_{a_2 b_2} \cdots R_{a_n b_n} E^{b_1 b_2 \cdots b_n}. \label{ETransformation}
\end{eqnarray}
for any $R \in \mathrm{SO}(N)_\mathrm{D}$ with $\det(R) = 1$.

If a number $i$ appears in the indices $\{a_1, a_2, \cdots, a_n\}$ for odd times, then $E^{a_1 a_2 \cdots a_n} = 0$.
This can be understood as follows.
Since $n<N$, we can always find an integer number $j$ ($1\leq j \leq N$) that doses not appear in $\{a_1, a_2, \cdots, a_n\}$ due to the pigeonhole principle.
Construct a diagonal matrix $R \in \mathrm{SO}(N)_\mathrm{D}$ with the elements $R_{ii}$ and $R_{jj}$ being $-1$ but the rest diagonal elements being $1$.
Applying Eq.~\eqref{ETransformation}, the $R_{ii} = -1$ appears for odd times in the right hand side and collectively contributes a $-1$ factor, leading to $E^{a_1 a_2 \cdots a_n} = - E^{a_1 a_2 \cdots a_n}$, and hence $E^{a_1 a_2 \cdots a_n}$ must vanish.

Therefore, any integer number $i$ ($1\leq i \leq N$) can only appear for even times in the indices of a nonzero $E^{a_1 a_2 \cdots a_n}$.
This implies that a nonzero $E^{a_1 a_2 \cdots a_n}$ tensor must have an even rank, i.e., an even $n$.

If $N$ is odd,  one can adopt $P_\mathrm{D} = \operatorname{diag}(-1, -1, \cdots, -1)$.
Applying this on $E^{a_1 a_2 \cdots a_n} V_1^{a_1} V_2^{a_2} \cdots V_n^{a_n}$ introduces a $(-1)^n=1$ factor because $n$ is even. Thus, $E^{a_1 a_2 \cdots a_n} V_1^{a_1} V_2^{a_2} \cdots V_n^{a_n}$ remains unchanged and is a $P_\mathrm{D}$-even scalar, contradicting our assumption.

If $N$ is even,  usually $P_\mathrm{D} = \operatorname{diag}(-1, 1, 1, \cdots, 1)$ is adopted without loss of generality. Under the $P_\mathrm{D}$ transformation, $V_i^1 \to - V_i^1$ and $V_i^j \to V_i^j$ ($j\neq 1$). Since $1$ must appear for even times in the indices of a nonzero $E^{a_1 a_2 \cdots a_n}$, we can still prove that $E^{a_1 a_2 \cdots a_n} V_1^{a_1} V_2^{a_2} \cdots V_n^{a_n}$ is $P_\mathrm{D}$-even.
This again contradicts with the assumption.

In summary, there is no way to construct a $P_\mathrm{D}$-odd scalar by contracting $n$ vectors for $n<N$,  so the generic scalar potential constructed from the $N-1$ Higgs multiplets $\Phi_i$ has an accidental global $\mathrm{O}(N)_\mathrm{D}$ symmetry.

If all $\Phi_i$ develop nonzero VEVs $\langle \Phi_i \rangle$, which are linearly independent, then the global $\mathrm{O}(N)_\mathrm{D}$ symmetry is spontaneously broken completely.  We are now going to seek for the remaining $Z'_2$ symmetry.  Similarly with what we have done in subsection~\ref{subsec:spon_sym_break}, we can always rotate the axes such that all the first VEV components $\langle \Phi_i^1 \rangle = 0$.
In fact, this VEV configuration is preserved under the $P'_\mathrm{D} = \operatorname{diag}(-1, 1, 1, \cdots, 1)$ transformation,  which is actually the $Z'_2$ symmetry we want.
Expanding the Higgs fields around the VEVs, $\Phi_i = \left<\Phi_i\right> + \phi_i$, we find that all $\phi_i^1$ ($i=1,2,\cdots, N-1$) are $P'_\mathrm{D}$-odd, while the rest $\phi_i^a$ ($a=2,\cdots, N$) are $P'_\mathrm{D}$-even.

There are $N(N-1)/2$ gauge bosons in such a $\mathrm{SO}(N)_\mathrm{D}$ gauge model. We denote the gauge fields as $A^{a b}_{\mu}$ ($a,b=1,2,\cdots, N$), where $A^{a b}_{\mu} = -A^{b a}_{\mu}$.
The trilinear gauge interaction terms are proportional to $A^{a b}_{\mu} \Phi_i^a \partial^{\mu} \Phi_i^b$.
Taking the VEVs, we have $A^{a b}_{\mu} \left<\Phi_i^a\right> \partial^{\mu} \Phi_i^b$, which only vanishes for $a=1$.
Therefore, the $N-1$ gauge fields $A_{\mu}^{a 1}$ ($a > 1$) are $P'_\mathrm{D}$-odd, while all the other $A_{\mu}^{a b}$ ($a \neq b$) are $P'_\mathrm{D}$-even.
The number of the $P'_\mathrm{D}$-odd gauge bosons is equal to the number of the $P'_\mathrm{D}$-odd scalar bosons $\phi_i^1$, implying that all these scalar bosons become the Goldstone bosons eaten by the $P'_\mathrm{D}$-odd gauge bosons.
The lightest mass eigenstate of the $P'_\mathrm{D}$-odd gauge bosons serves as a DM candidate.
In this way, our model setup has been generalized to the $\mathrm{SO}(N)_\mathrm{D}$ cases.

\section{Conclusions and future prospect}
\label{sec:conclusion}

In this paper, we have discussed a vector DM model  based on a dark $\NGG$ gauge theory.
Two real $\NGG$ dark Higgs triplets are introduced to completely break the $\NGG$ gauge symmetry, leading to three massive dark gauge bosons $A^1$, $A^2$, and $A^3$ with a split spectrum.
An accidental $Z'_2$ symmetry remains after the spontaneous symmetry breaking.
Under this $Z'_2$ symmetry, $A^1$ is even, while both $A^2$ and $A^3$ are odd.
Since $A^2$ is generally lighter than $A^3$, it becomes a stable vector DM candidate.
In the unitary gauge, the physical degrees of freedom in the dark Higgs triplets and the $\mathrm{SU(2)_L}$ Higgs doublet form four Higgs bosons $h_0$, $h_1$, $h_2$, and $h_3$, where $h_0$ acts as the 125~GeV SM-like Higgs boson.
These Higgs bosons provide a portal to the SM particles for the dark gauge bosons.

We have randomly scanned the 14-dimensional parameter space, taking into account the constraints from current 125~GeV Higgs measurements, direct searches for exotic Higgs bosons, and electroweak precision measurements.
The DM relic density has been calculated, including the $A^2 A^3$ coannihilation effect.
We have found the parameter points predicting the observed relic density.
These parameter points have been further tested by the bounds from the XENON1T direct detection experiment and the Fermi-LAT indirect detection experiment.

Resonance and threshold effects in DM annihilation could lead to important differences between $\sv_{\mathrm{FO}}$ and $\sv_0$.
In the parameter points, we have found significant effects due to the $h_0$, $h_1$, and $h_2$ resonances for $m_{A^2} \sim m_{h_i}/2$, as well as a remarkable $h_1 h_1$ threshold effect for $m_{A^2} \sim m_{h_1}$.
Because of these effects, $\sigma_N^\mathrm{SI}$ and $\sv_0$ for some parameter points could be sufficiently small, evading the current direct and indirect detection constraints.

There are numerous parameter points remaining after all the above constraints are considered.
The future LZ direct detection experiment is expected to test some of them.
Moreover, the interactions between the dark Higgs triplets and the $\mathrm{SU(2)_L}$ Higgs doublet induce mixings among the Higgs bosons.
As a result, the couplings of the SM-like Higgs boson $h_0$ generally deviate from the SM couplings.
Future Higgs precision measurements at $e^+e^-$ colliders, such as CECP~\cite{CEPCStudyGroup:2018ghi}, FCC-ee~\cite{Abada:2019lih}, and ILC~\cite{Baer:2013cma}, will provide further tests on this model.

We have proved that the way to construct this model can be generalized to the general $\mathrm{SO}(N)_\mathrm{D}$ cases.
For a $\mathrm{SO}(N)_\mathrm{D}$ gauge model, $N-1$ real Higgs multiplets in the $\mathrm{SO}(N)_\mathrm{D}$ fundamental representation can be introduced to break the gauge symmetry, with a remaining $Z'_2$ symmetry ensuring the stability of a dark gauge boson.
Thus, more vector DM models can be similarly constructed.

\begin{acknowledgments}

We thank Gao-Liang Zhou for helpful discussions.
This work is supported in part by the National Natural Science Foundation of China under Grants No.~11805288, No.~11875327, No.~11905300, and No.~12005312, the China Postdoctoral Science Foundation
under Grant No.~2018M643282,
the Fundamental Research Funds for the Central Universities,
and the Sun Yat-Sen University Science Foundation.

\end{acknowledgments}

\bibliographystyle{utphys}
\bibliography{reference}

\end{document}